%% file: NGC5128.tex
\documentclass[11pt,a4paper]{article}
\usepackage[left=15mm, top=10mm, right=15mm, bottom=20mm, nohead=true, nofoot=true]{geometry}

\usepackage[utf8]{inputenc}
\usepackage{authblk}
\usepackage{indentfirst}
\usepackage{misccorr}
\usepackage{graphicx}
\usepackage{amsmath}
\usepackage{amssymb}
\usepackage{multicol}
\usepackage{bm}
\usepackage{longtable}
\usepackage{pdflscape}
\usepackage{epstopdf}
\usepackage{multirow}
\usepackage{adjustbox}
\usepackage{amsfonts}
\usepackage{ifthen}
\usepackage{fancyhdr}
\usepackage{setspace}
\usepackage{xcolor}
\usepackage[round,authoryear,comma,sort&compress,numbers]{natbib}
\usepackage[toc,title,page]{appendix}
\usepackage[hidelinks]{hyperref}
\usepackage{url}

\hypersetup{
    colorlinks=true,
    linkcolor=green,
    citecolor=blue,
    filecolor=magenta,
    urlcolor=cyan,
    pdftitle={Sharelatex Example},
    pdfpagemode=FullScreen,
}

\fancypagestyle{plain}{\chead{\texttt{\textcolor{brown}{Communications of BAO, Vol. ?, Issue ?, 20??, pp. ???-???}}}}
\title{\textbf{Deciphering Galactic Halos: A Detailed Review of Star Formation in NGC 5128 (Cen A)}}

\author[1]{M. Abdollahi \thanks{m.abdollahi@ipm.ir, Corresponding author}}
\author[1]{S. T. Aghdam}
\author[1]{A. Javadi \thanks{atefeh@ipm.ir}}
\author[2,3]{S. A. Hashemi}
\author[4]{J. Th. van Loon}
\author[1]{H. Khosroshahi}
\author[5]{R. Hamedani Golshan}
\author[1,6,7]{E. Saremi}
\author[8]{M. Saberi}
\affil[1]{\scriptsize School of Astronomy, Institute for Research in Fundamental Sciences (IPM), Tehran, 19568-36613, Iran}
\affil[2]{\scriptsize Department of Physics, Sharif University of Technology, Tehran, 11155-9161, Iran}
\affil[3]{\scriptsize Department of Physics and Astronomy, University of California Riverside, CA 92521, The USA}
\affil[4]{\scriptsize Lennard-Jones Laboratories, Keele University, ST5 5BG, UK}
\affil[5]{\scriptsize I. Physikalisches Institut, Universität zu Köln, Zülpicher Straße 77, 50937 Cologne, Germany}
\affil[6]{\scriptsize Instituto de Astrof{\`i}sica de Canarias, C/ V{\`i}a L{\`a}ctea s/n, 38205 La Laguna, Tenerife, Spain}
\affil[7]{\scriptsize Departamento de Astrof{\`i}sica, Universidad de La Laguna, 38205 La Laguna, Tenerife, Spain}
\affil[8]{\scriptsize Rosseland Centre for Solar Physics, University of Oslo, P.O. Box 1029, Blindern, NO-0315, Oslo, Norway}

\input jr.tex

\begin{document}
\pagestyle{empty}
\newpage
\pagestyle{fancy}
\label{firstpage}
\date{}
\maketitle
\begin{abstract}
NGC 5128 (Centaurus A), the closest giant elliptical galaxy outside the Local Group to the Milky Way, is one of the brightest extragalactic radio sources. It is distinguished by a prominent dust lane and powerful jets, driven by a supermassive black hole at its core. Using previously identified long-period variable (LPV) stars from the literature, this study aims to reconstruct the star formation history (SFH) of two distinct regions in the halo of NGC 5128. These regions reveal remarkably similar SFHs, despite being located about 28 kpc apart on opposite sides of the galaxy’s center. In Field 1, star formation rates (SFRs) show notable increases at approximately 800 Myr and 3.8 Gyr ago. Field 2 exhibits similar peaks at these times, along with an additional rise around 6.3 Gyr ago. The increase in SFR around 800 Myr ago is consistent with earlier research suggesting a merger event. Since no LPV catalog exists for the central region of NGC 5128, we focused our investigation on its outer regions, which has provided new insights into the complex evolutionary history of this cornerstone galaxy. The SFH traced by LPVs supports a scenario in which multiple events of nuclear activity have triggered episodic, jet-induced star formation.
\end{abstract}

\section{Introduction}

Our position within the Local Group (LG) provides an excellent opportunity to study the resolved stellar populations of spiral galaxies, enhancing our understanding of their formation and evolution. However, because the LG lacks a giant elliptical (GE) galaxy, our knowledge of such systems relies on observations of the nearest elliptical galaxies in neighboring groups. NGC 5128 (Centaurus A), situated 3.8 Mpc away ($\mu = 27.87 \pm 0.16 $ mag; \citealp{Rejkuba2004a}, and $E(B-V) = 0.15 \pm 0.05$ mag; \citealp{Rejkuba2004b}), presents a rare opportunity to study a nearby GE galaxy in details (\citealp{harris1999}; \citealp{charmandaris2000}; \citealp{Rejkuba2004b}, \citeyear{Rejkuba2005}), as it resides within the Centaurus galaxy group (\citealp{karachentsev2005}).

NGC 5128 is considered a post-merger galaxy (\citealp{Peng2002}) and is one of the few halos that has been resolved into individual stars (\citealp{Rejkuba2011}). The stellar populations in NGC 5128’s halo serve as crucial indicators of its star formation history, revealing evidence of past interactions and mergers.

The active galactic nucleus (AGN) at the center of NGC 5128 drives powerful radio jets, providing the closest example of this galactic outflows (\citealp{crockett2012}). How AGN activity influences star formation and the evolution of its host galaxy remains a crucial but unresolved question in galaxy formation and evolution theory (\citealp{ciotti1997}; \citealp{silk1998}, \citeyear{silk2005}; \citealp{binney2004}; \citealp{springel2005}; \citealp{sijacki2007}; \citealp{schawinski2007}).

The star formation history is essential for understanding galaxy formation and evolution. In resolved galaxies, SFH is usually derived from color–magnitude diagrams (CMDs) of individual stars, revealing signatures of various stellar populations (e.g., \citealp{tolstoy1996}; \citealp{holtzman1999}; \citealp{olsen1999}; \citealp{dolphin2002},\citealp{javadi2011b}). However, this analysis is typically limited to a few dozen galaxies, mostly within our Local Group, due to observational constraint.

This work aims to determine the SFH of two small fields in the halo of NGC 5128 by using long-period variable stars as tracers, to explore the connection between the halo's SFH and its merger history.

\section{Data}

\begin{figure}
\centering
\includegraphics[width=1\textwidth]{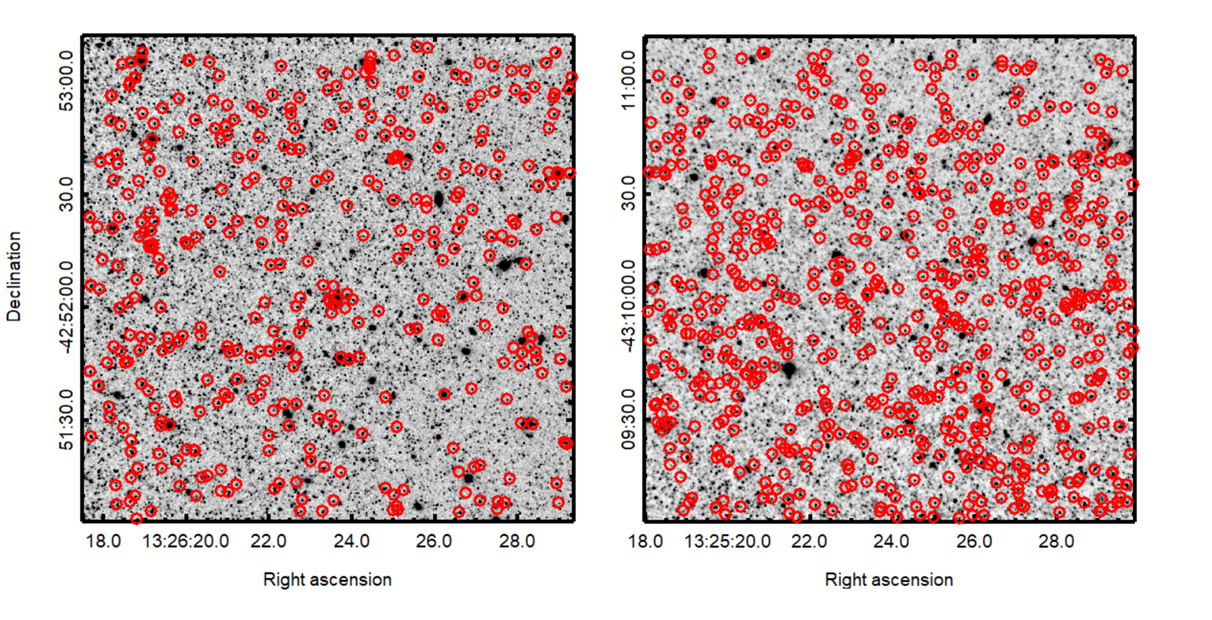}
\caption{The images of the fields studied here (each approximately $2.3 \times 2.3$ arcminutes) are shown. Red circles indicate the locations of the selected LPV stars. (Left panel) The northwestern field is centered at $\alpha = 13^{h}\ 26^{m}\ 23\rlap{.}^{s}5$, $\delta = -42^{\circ}\ 52^{\prime}\ 0^{\prime\prime}$, on the eminent northeastern part of the halo, at a distance of $\sim 17^{\prime}$ ($\sim 18.8$ kpc) from the center of the galaxy presented by \citealp{ma1998}, with dimensions of $2\rlap{.}^{\prime}28 \times 2\rlap{.}^{\prime}30$ ($5.7$ kpc$^2$). (Right panel) The southern field is centered at $\alpha = 13^{h}\ 25^{m}\ 26^{s}$, $\delta = -43^{\circ}\ 10^{\prime}\ 0^{\prime\prime}$, at a distance of $\sim 9^{\prime}$ ($\sim 9.9$ kpc) from the center, with dimensions of $2\rlap{.}^{\prime}25 \times 2\rlap{.}^{\prime}31$ ($5.7$ kpc$^2$).
}
\label{fig: fileds}
\end{figure}

We utilized near-infrared data published by \citealp{Rejkuba2001} and analyzed by \citealp{Rejkuba2003a}. This catalog was obtained using the ISAAC instrument on the ESO Paranal UT1 Antu 8.2 m Very Large Telescope, covering two distinct fields in the northwestern and southern regions of the halo of NGC 5128  (referred to as Field 1 and Field 2, respectively), as shown in Figure \ref{fig: fileds}.

\cite{Rejkuba2004b} identified 15,574 and 18,098 stars in Fields 1 and 2, respectively. However, our focus is on long-period variable stars with periods exceeding 70 days, which form the basis of a method for constructing the star formation history as developed by \cite{javadi2011a}.

\cite{Rejkuba2003a} identified LPV stars using multi-epoch photometry in the Ks band, along with single-epoch photometry in the Js and H bands, in two mentioned fields. However, based on the specified criteria, we selected 395 LPV stars in Field 1 and 671 in Field 2. The distribution of these selected LPV stars, marked by red circles, is displayed in Figure \ref{fig: fileds}.

\section{Method}

As mentioned in the previous section, \citet{javadi2011a} developed a method for calculating star formation histories based on long-period variable stars. During the LPV phase, stars reach their peak luminosity, providing a valuable opportunity to establish a strong correlation between birth mass and this peak. Based on this correlation, relations—including the birth mass-luminosity, age-mass, and pulsation duration relations—can be derived for each metallicity from the Padova stellar evolutionary models \citep{Marigo2017}. By inputting the stars' magnitudes into these relations, we can determine the components of an equation that statistically reconstructs the star formation history of galaxies based on the initial mass function.

\begin{equation}
\xi(t) = \frac{dn^\prime(t)}{\delta t}\ \frac{\int_{\rm min}^{\rm max}f_{\rm IMF}(m)m\ dm}
{\int_{m(t)}^{m(t+dt)}f_{\rm IMF}(m)\ dm},
\label{eq:eq1}
\end{equation}

where $m$ is birth mass and $f_{\rm IMF}(m)$ is Kroupa initial mass function (IMF) \citep{Kroupa2001}.

\begin{figure}
\centering
  \includegraphics [width=0.48\textwidth] {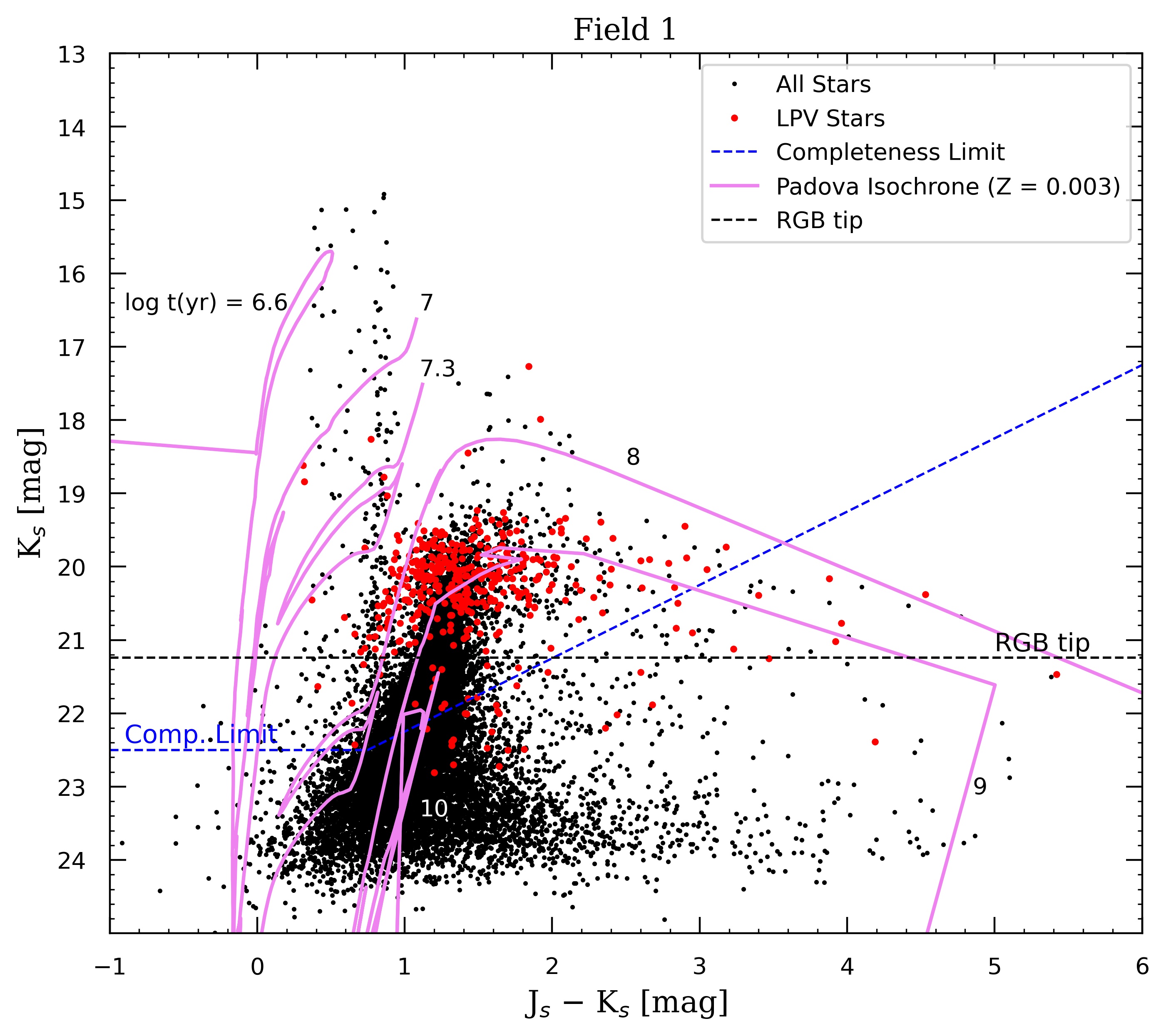}
  \includegraphics [width=0.48\textwidth] {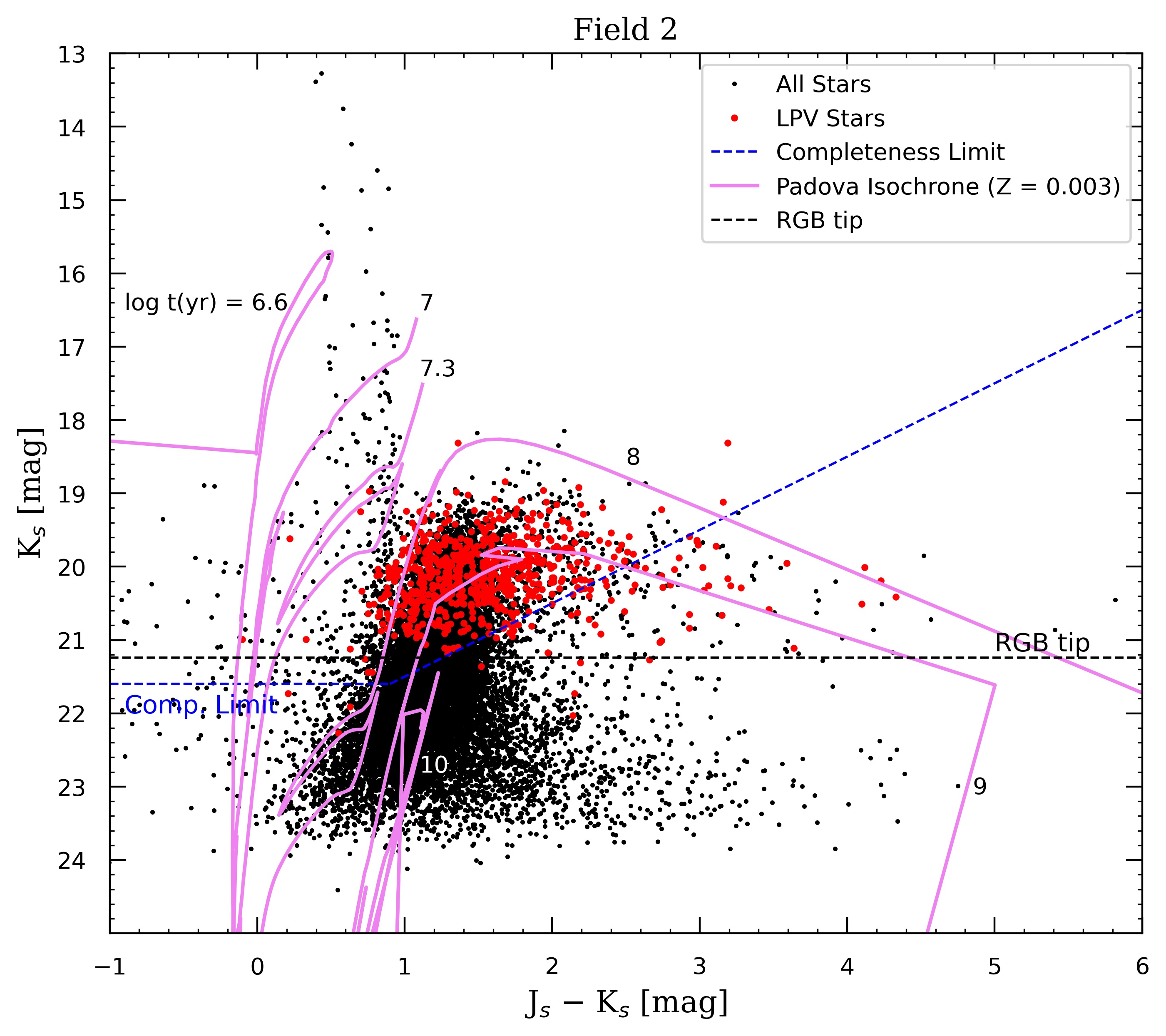}
  \caption{The $K_s$ vs. $J_s - K_s$ color-magnitude diagram (CMD) shows stars with at least three $K_s$-band detections (black dots) and long-period variable (LPV) stars (red dots) in Field 1 (left) and Field 2 (right). The black and blue dotted lines represent the tip of the red giant branch (RGB) and the completeness limit magnitudes, respectively, for each field (\citealp{Rejkuba2003a}). The purple lines indicate theoretical stellar isochrones for a metallicity of $Z = 0.003$ across six different ages (\citealp{Marigo2017}).}
  \label{fig: CMD}
\end{figure}

\begin{table}
\centering
\caption{The age-metallicity relation investigated by \cite{woodley2010b} and \cite{yi2004} assuming Z$_\odot$ = 0.0198 (\citealp{Rejkuba2011}).}

\begin{tabular}{lclc} \hline
\multicolumn{2}{c}{\citealp{woodley2010b}} & \multicolumn{2}{c}{\citealp{yi2004}} \\ \hline
Age range (Gyr) & Z       & Age range (Gyr) & Z      \\ \hline
age $\geq$ 12   & 0.001   & age $\geq$ 10   & 0.0003 \\
8 $\leq$ age \textless 12 & 0.003 & 6 $\leq$ age \textless 10 & 0.001  \\
6.5 $\leq$ age \textless 8 & 0.006 & 4 $\leq$ age \textless 6  & 0.003  \\
5.5 $\leq$ age \textless 6.5 & 0.008 & 3 $\leq$ age \textless 4  & 0.010  \\
3 $\leq$ age \textless 5.5 & 0.010 & 2 $\leq$ age \textless 3  & 0.020  \\
2 $\leq$ age \textless 3   & 0.020 & age \textless 2           & 0.039  \\
age \textless 2            & 0.030  &                   &        \\ \hline
\end{tabular}
\label{tab: Age-metallicity}
\end{table}

As shown in Figure \ref{fig: CMD}, we expect all LPV stars to be located near the peaks of the isochrones derived from the Padova stellar evolutionary models. However, some stars appear spread into redder regions due to surrounding dust. To correct for dust effects, these stars need to be shifted back to the isochrone peaks before their magnitudes are used in the mentioned statistical equations.

This method, which has since been applied in a variety of studies (\citealp{javadi2011a}, \citeyear{javadi2011b}, \citeyear{javadi2017}; \citealp{Rezaei2014}; \citealp{golshan2017}; \citealp{hashemi2019}; \citealp{navabi2021}; \citealp{saremi2021}; \citealp{parto2023}; \citealp{abdollahi2023};\citealp{Aghdam2024NGC5128}; \citealp{Khatamsaz2024}), considers a range of metallicities to define the epochs of star formation for each case and enables comparisons among them.

However, we go a step further by incorporating the age-metallicity relation, which provides an opportunity to address uncertainties in the metallicity. To elaborate, this approach allows us to determine the metallicity of the galaxy across each age range, as presented in Table \ref{tab: Age-metallicity}.

\begin{figure}
\centering
  \includegraphics [width=0.48\textwidth] {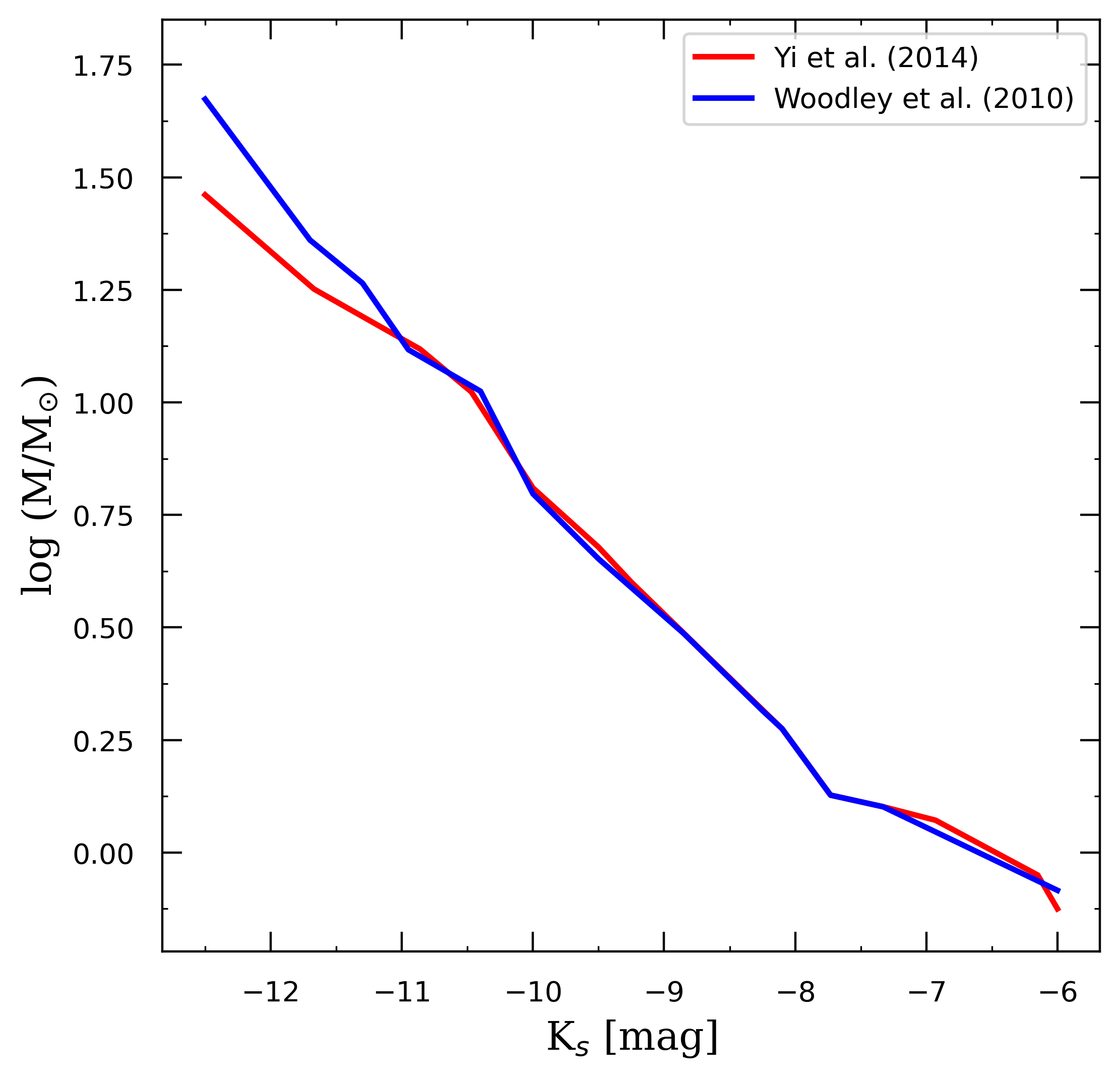}
  \includegraphics [width=0.48\textwidth] {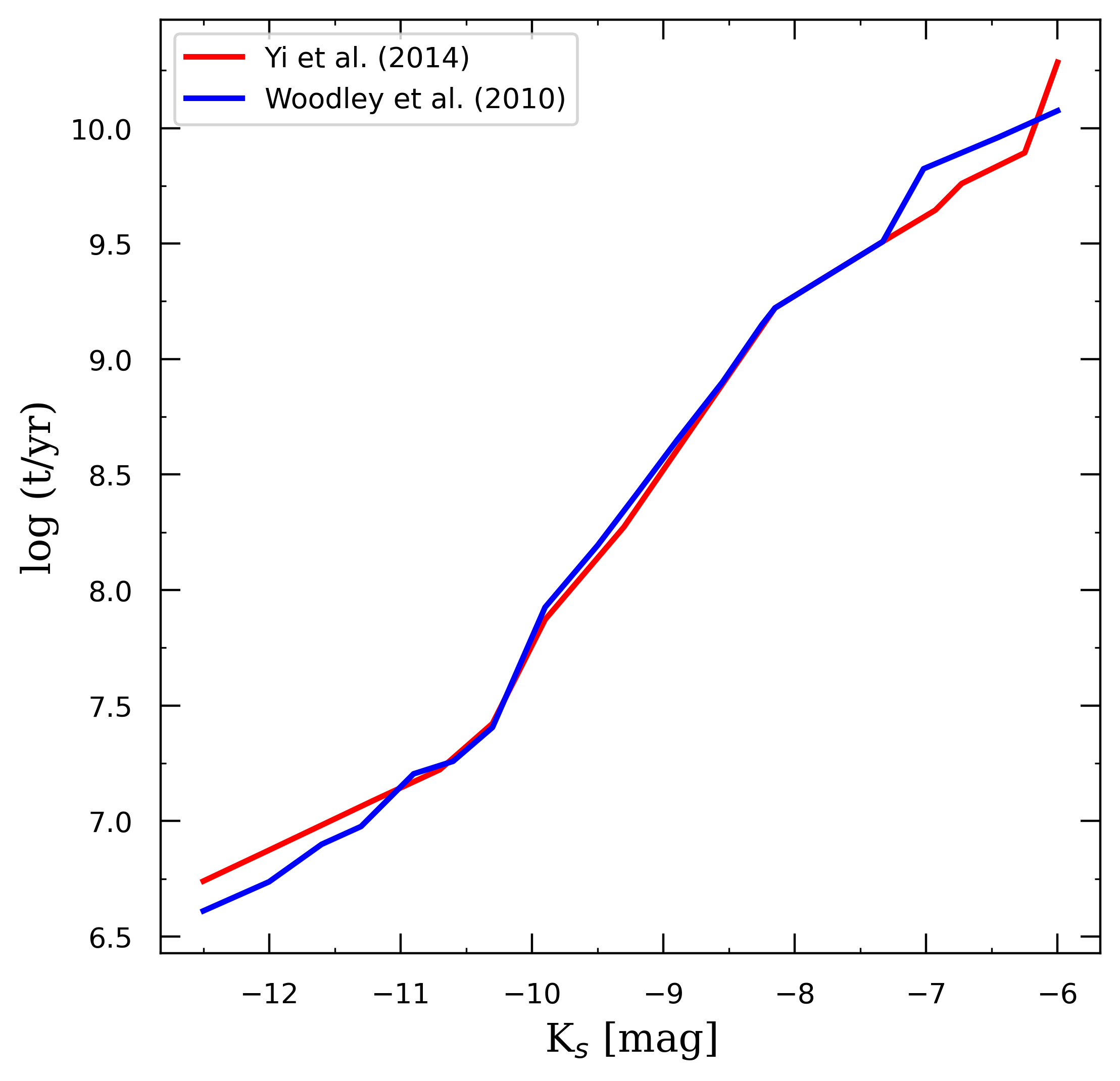}
  \caption{The mass–luminosity (left panel) and age–luminosity (right panel) relations, considering the AMRs determine by \cite{yi2004} (red line) and \cite{woodley2010b} (blue line).}
  \label{fig: AMR_models}
\end{figure}

We redefined the mass-luminosity and age-mass relations based on the age-metallicity relation (AMR), resulting in a single, unified relation that encompasses the full range of metallicities experienced by the galaxy. Figure \ref{fig: AMR_models} illustrates these relations for the models determined by \cite{yi2004} and \cite{woodley2010b}. As shown, the trends in these diagrams are similar, with only slight differences at the beginning and end of the intervals. Using the AMR approach, we can now overcome the metallicity degeneracy and establish Equation \ref{eq:eq1} to derive the SFH for the two separate fields.

\section{Results and Discussion}

\begin{figure}[ht]
\centering
  \includegraphics [width=0.48\textwidth] {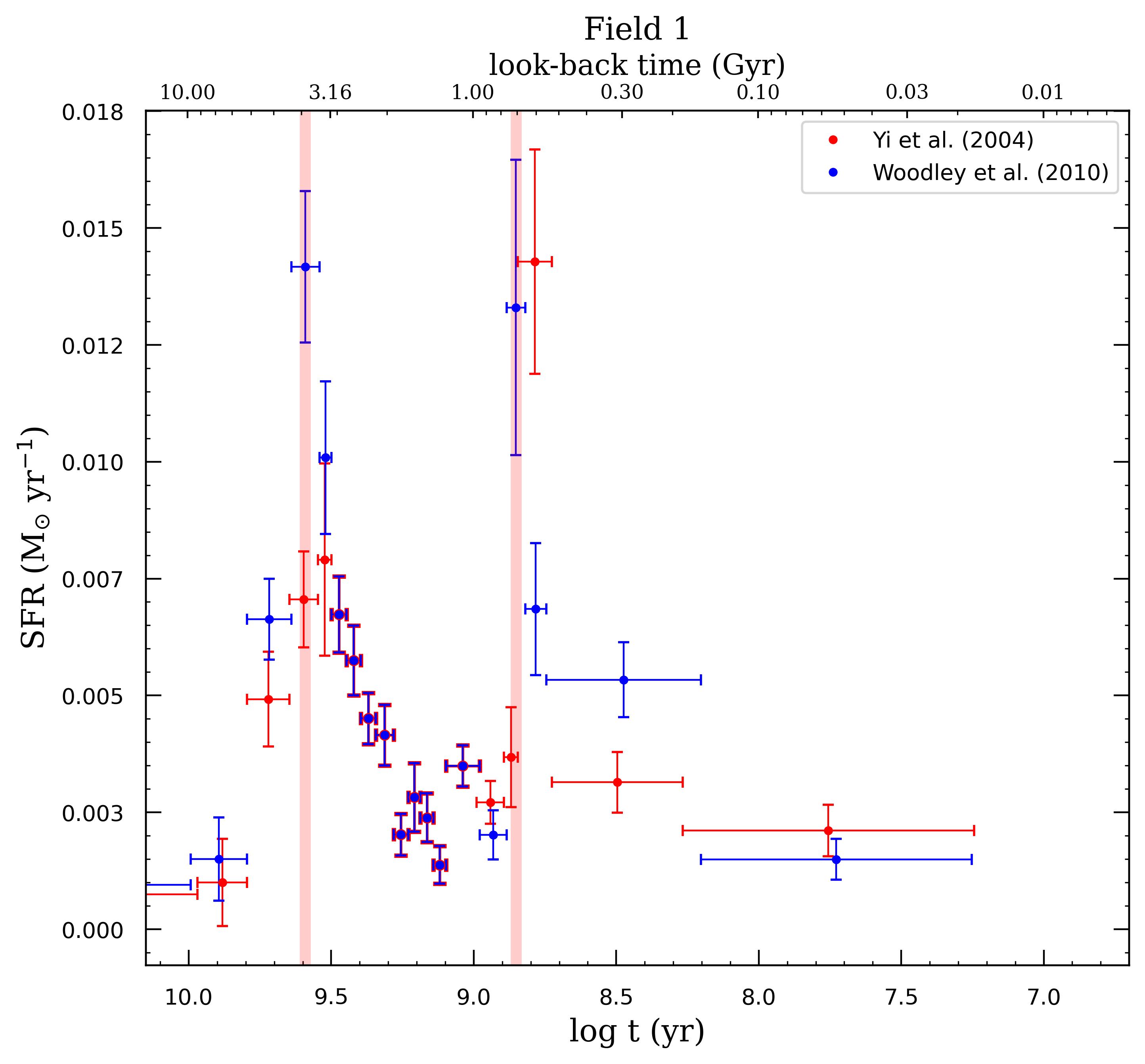}
  \includegraphics [width=0.48\textwidth] {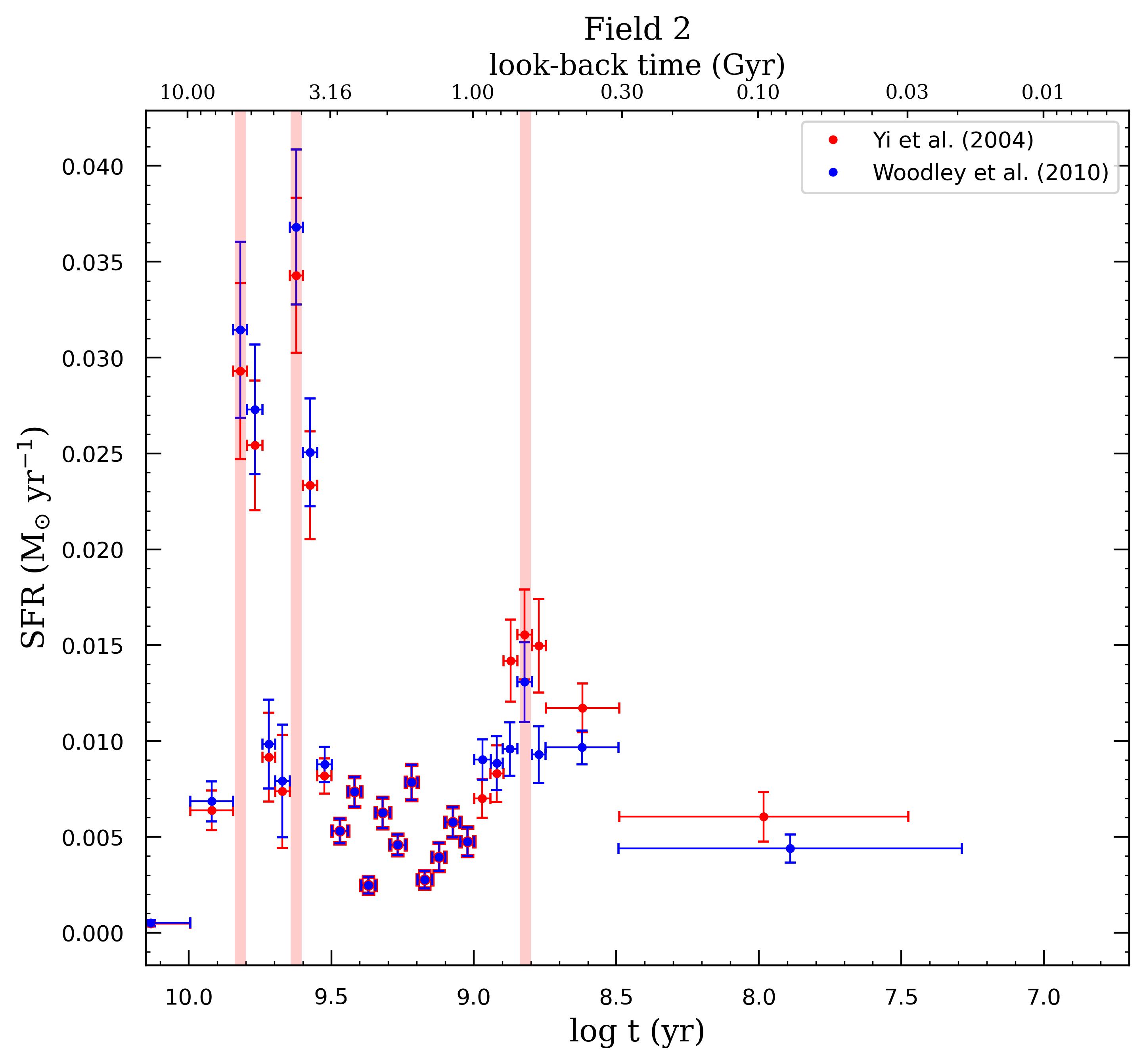}
  \caption{The SFH derived using the age-metallicity relations (AMRs) from \cite{yi2004} (red markers) and \cite{woodley2010b} (blue markers) is shown for Field 1 (left panel) and Field 2 (right panel). Highlighted regions indicate the peaks of star formation during major epochs.}
  \label{fig: AMR_results}
\end{figure}

In our analysis of the SFH of NGC 5128, illustrated in Figure \ref{fig: AMR_results}, Field 1 exhibited two major star formation epochs: one approximately 3.6 billion years ago and another around 800 million years ago. Field 2, however, showed three significant epochs, occurring roughly 6.3 billion, 3.6 billion, and 800 million years ago. The latter two epochs align with those in Field 1, suggesting a connection between these distant regions, despite being separated by 28 kpc on opposite sides of the galaxy. This similarity in SFH across such large distances implies shared galactic-scale influences.

Comparing our results with previous studies reveals further insights into the history of NGC 5128. Evidences such as the galaxy’s unusual structure (\citealp{graham1979}), optical and neutral hydrogen shells (\citealp{malin1983}; \citealp{Peng2002}; \citealp{schiminovich1994}), and ongoing star formation suggest a history of mergers and interactions (\citealp{Rejkuba2001}). Specifically, literature indicates a major merger with a smaller, gas-rich galaxy around 1 Gyr ago (\citealp{malin1983}; \citealp{sparke1996}), which aligns with our observed peak in star formation around the previous 800 Myr. Additionally, a possible minor merger could explain the secondary star formation peak, indicating this galaxy has undergone multiple interactions over time.

Furthermore, NGC 5128’s central supermassive black hole has likely driven active galactic nucleus (AGN) activity, fueled by abundant cold gas. Such activity, as noted in studies by \citealp{Saxton2001} and \citealp{Hardcastle2009}, has been linked to increased the star formation rate. The age estimates for the nuclear molecular layer ($\sim$150 Myr) and inner lobes ($\sim$30 Myr) suggest that the AGN was active well before these epochs, potentially contributing to the sustained rise in star formation observed up to around 800 Myr ago.

\section{Conclusion}

In this study, we investigated the star formation history (SFH) of NGC 5128, the nearest giant elliptical galaxy to the Local Group, by employing a statistical approach based on long-period variable (LPV) stars. Our analysis focused on two distinct regions within the halo of this galaxy, providing insights into the star formation activity in the outer regions of this unique system. Key findings include:

\begin{itemize}
\item A novel method based on the age-metallicity relation of LPV stars was introduced and employed to investigate the SFH of NGC 5128.
\item Despite their significant separation, both studied regions exhibited a similar SFH, suggesting equivalent evolutionary influences across the galaxy.
\item Evidence of peaks in star formation supports the occurrence of a past collision or merger in NGC 5128.
\item Our findings indicate that active galactic nucleus (AGN) activity can enhance the star formation rate in galaxies.
\end{itemize}

\scriptsize
\bibliographystyle{ComBAO}
\nocite{*}
\bibliography{NGC5128}

\end{document}

%% file: jr.tex
\def\apj{Astrophys.~J. }

\def\aap{Astron.~Astrophys. }

\def\apjl{Astrophys.~J.~Lett. }

\def\aj{Astron.~J. }
\def\mnras{Mon.~Not.~R.~Astron.~Soc. }

